\documentclass{ieeetj}
\usepackage{cite}
\usepackage{amsmath,amssymb,amsfonts}
\usepackage{algorithmic}
\usepackage{graphicx,color}
\usepackage{textcomp}
\usepackage{xcolor}
\usepackage{hyperref}
\hypersetup{hidelinks=true}
\usepackage{algorithm,algorithmic}
\usepackage{tikz}
\usepackage{pgfplots}
\pgfplotsset{compat=1.17}
\usetikzlibrary{shapes, positioning, arrows.meta, shapes.geometric}
\usepackage{pifont} 
\usepackage{booktabs}
\usepackage{tabularx}
\usepackage{multirow}
\usepackage{array}
\usepackage{float}
\usepackage{comment}
\usepackage{placeins}
\usepackage[T1]{fontenc}

\def\BibTeX{{\rm B\kern-.05em{\sc i\kern-.025em b}\kern-.08em
    T\kern-.1667em\lower.7ex\hbox{E}\kern-.125emX}}
\AtBeginDocument{\definecolor{tmlcncolor}{cmyk}{0.93,0.59,0.15,0.02}\definecolor{NavyBlue}{RGB}{0,86,125}}

\definecolor{brightblue}{RGB}{0, 102, 204}
\definecolor{brightred}{RGB}{204, 0, 0}
\definecolor{brightgreen}{RGB}{0, 153, 0}

\def\OJlogo{\vspace{-4pt}$<$Society logo(s) and publication title will appear here.$>$}
\def\seclogo{\vspace{10pt}$<$Society logo(s) and publication title will appear here.$>$}

\def\authorrefmark#1{\ensuremath{^{\textbf{#1}}}}

\makeatletter
\renewcommand\section{\@startsection{section}{1}{\z@}%
  {-2.5ex \@plus -1ex \@minus -.2ex}
  {1.3ex \@plus .2ex}
  {\normalfont\normalsize\bfseries}}

\renewcommand\subsection{\@startsection{subsection}{2}{\z@}%
  {-2.0ex \@plus -1ex \@minus -.2ex}
  {0.8ex \@plus .2ex}
  {\normalfont\normalsize\bfseries}}

\renewcommand\subsubsection{\@startsection{subsubsection}{3}{\z@}%
  {-1.5ex \@plus -1ex \@minus -.2ex}
  {0.5ex \@plus .2ex}
  {\normalfont\normalsize\bfseries}}
\makeatother

\begin{document}
\receiveddate{XX Month, XXXX}
\reviseddate{XX Month, XXXX}
\accepteddate{XX Month, XXXX}
\publisheddate{XX Month, XXXX}
\currentdate{XX Month, XXXX}
\doiinfo{XXXX.2024.XXXXXXX}

\markboth{}{}

\title{SkyHOST: A Unified Architecture for Cross-Cloud Hybrid Object and Stream Transfer}

\author{MUHAMMAD ARSLAN TARIQ\authorrefmark{1}, GRÉGOIRE DANOY\authorrefmark{1,2}  (Member, IEEE), AND PASCAL BOUVRY\authorrefmark{1,2} (Member, IEEE)}

\affil{SnT, University of Luxembourg, Luxembourg}
\affil{FSTM/DCS, University of Luxembourg, Luxembourg}
\corresp{Corresponding author: Muhammad Arslan Tariq (email: arslan.tariq@uni.lu).}
\authornote{This work is partially funded by the SnT-LuxProvide partnership on bridging clouds and supercomputers and by the Fonds National de la Recherche Luxembourg (FNR) POLLUX program under the SERENITY Project (ref.C22/IS/17395419).}

\begin{abstract}
Cloud and big data workloads are increasingly distributing data across multiple cloud providers and regions for rapid decision-making and analytics. Traditional transfer tools are typically specialized for a single paradigm, either stream replication or bulk transfer. This specialization forces users to deploy and manage separate systems with different configurations for each transfer pattern. This paper presents SkyHOST (Hybrid Object and Stream Transfer), a unified data movement architecture built upon the Skyplane framework to bridge the gap between bulk object transfer and streaming workloads through a single control plane and CLI. SkyHOST manages URI-based routing to automatically select the appropriate transfer mechanism, supporting both structured data for record-level ingestion and chunk-based transfer for large binary objects. We demonstrate, through an environmental monitoring use case and empirical evaluation, that SkyHOST provides operational simplicity by consolidating heterogeneous data movement patterns under a single control plane while achieving competitive throughput for cross-region transfers.
\end{abstract}


\maketitle

\section{INTRODUCTION}

Efficient data movement across heterogeneous cloud environments has become a critical challenge as data volumes and cross-cloud adoption continue to grow \cite{data_intensive}. Modern organizations increasingly rely on hybrid cloud deployments to support workloads ranging from large-scale binary object transfers to continuous data stream replication. However, transferring data across multiple cloud platforms remains complex, due to the coexistence of heterogeneous data sources and a fragmented ecosystem of transfer tools. Therefore, users are forced to deploy and maintain multiple specialized tools, each tailored to specific workload characteristics. When managing hybrid workloads at scale, ranging from terabytes to petabytes of data per day \cite{tbtopb}, this fragmentation introduces significant performance, scalability, and operational bottlenecks. These limitations expose a fundamental gap in the unified management of heterogeneous data sources across both industrial and scientific domains.

A unified architecture that integrates real-time data streams with historical bulk data provides substantial operational and analytical advantages. In healthcare, for example, systems must manage the bulk transfer of high-resolution medical images, such as tomography scans \cite{tomography}, while also ingesting low-latency IoT data streams for monitoring and diagnostics \cite{health1},\cite{health2},\cite{health3}. Environmental monitoring platforms and smart city infrastructures face similar requirements, combining satellite imagery with continuous sensor streams to enable real-time alerting and large-scale analytics \cite{IoT1},\cite{IoT2},\cite{smartagriculture},\cite{smartcities}. These systems support critical applications in climate monitoring, disaster response, medical care delivery, and agriculture monitoring, where architectural fragmentation and operational complexity can delay critical decision-making.  

However, addressing these hybrid requirements remains challenging due to the lack of unified support in existing data management tools. Current platforms specialize in either high-throughput, chunk-based bulk data transfer, such as Skyplane \cite{jain2023skyplane} and GridFTP \cite{globus}, or low-latency, record-aware data streaming like Apache Kafka \cite{kafka} and Amazon Kinesis \cite{amazonrecords}.

The challenge of unifying batch and streaming data processing has been recognized across multiple research domains. Foundational architectural patterns, like Lambda and Kappa, introduced hybrid processing models, while programming models like Apache Beam provided a unified API for both batch and streaming computations \cite{dataflow}. In the context of data ingestion, frameworks like Gobblin \cite{gobblin} and multi-cloud stream processing architectures like MC-BDP \cite{vergilio2023unified} provide unified processing capabilities. However, these existing approaches primarily focus on data processing rather than cross-cloud data movement, and no single system provides unified data movement across cloud boundaries while handling both bulk object transfer and stream replication.

To address this gap, we present SkyHOST, a unified data movement framework that efficiently supports both bulk object transfer and stream replication through a single control plane and command-line interface (CLI). Building upon Skyplane's high-throughput bulk transfer capabilities \cite{jain2023skyplane}, SkyHOST extends the system to enable stream-to-stream and object-to-stream transfers. The framework employs URI-based routing to automatically select appropriate operators (object store vs. stream), supports structured data formats (CSV, JSON) for record-level ingestion, and uses chunk-based transfer for large binary objects. This unified design allows a single CLI and control plane to dynamically adapt its transfer strategy based on source characteristics, data formats, and workload requirements. We validate SkyHOST through a comparative evaluation against specialized tools i.e., Confluent Kafka Replicator and S3 Source Connector demonstrating that it achieves competitive performance for bulk transfers and stream replication while significantly reducing operational complexity.

In this article, we introduce the following contributions:
\begin{enumerate}
\item We design and implement SkyHOST, a unified data movement architecture that extends Skyplane to support both object-to-stream and stream-to-stream transfers through URI-based routing and source-specific operator pipelines.

\item We develop and validate analytical performance models for both stream replication and bulk transfer, enabling parameter selection and performance prediction.

\item We demonstrate, through an environmental monitoring use case and empirical evaluation, that the proposed unified architecture streamlines heterogeneous data movement patterns (historical S3 datasets and Kafka streams) by providing operational simplicity and unified management within a single control plane. 
\end{enumerate}

The remainder of this paper is as follows: Section \ref{sec:related} reviews related work, identifies the research gap, and motivates our contribution. Section \ref{sec:Architecture} presents the SkyHOST architecture and its design principles. In Section \ref{sec:Performance}, we introduce the performance modeling, followed by Section \ref{sec:System}, where we discuss the SkyHOST implementation details. Section \ref{sec:Evaluation} presents the evaluation setup, use case evaluation, and model validation. Finally, in Section \ref{sec:Conclusion}, we conclude the paper and suggest future directions.

\begin{table*}[t!]
\centering
\caption{Comparative Analysis of Data Movement Systems}
\label{tab:taxonomy_comparison}
\resizebox{\textwidth}{!}{%
\renewcommand{\arraystretch}{1.2}
\begin{tabular}{@{}lccccc@{}}
\toprule
\textbf{System} & \textbf{Data Model} & \textbf{Optimized For} & \textbf{Cross-Cloud Transfer} & \textbf{Deployment} & \textbf{Availability} \\
\midrule
\multicolumn{6}{c}{\textit{High-Throughput Bulk Transfer Systems}} \\
Skyplane \cite{jain2023skyplane} & File/Object Transfer & Bulk Throughput & Yes (Optimized Routing) & Multi-Cloud & Open Source \\
Blaze \cite{Blaze} & File/Object Transfer & Bulk Throughput & Yes (Single-Agent) & Multi-Cloud & Open Source \\
Rucio \cite{rucio} & File/Dataset Transfer & Bulk Throughput & Yes (Multi-Site) & On-Prem, Cloud & Open Source \\
GridFTP \cite{globus} & File Transfer & Bulk Throughput & Yes (Multi-Site) & On-Prem, Cloud & Open Source \\
\midrule
\multicolumn{6}{c}{\textit{Low-Latency Streaming Systems}} \\

Apache Kafka \cite{kafka} & Record/Stream & Real-time Latency & Yes (MirrorMaker 2) & On-Prem, Cloud & Open Source \\
Apache Pulsar \cite{pulsar} & Record/Stream & Real-time Latency & Yes (Geo-Replication) & On-Prem, Cloud & Open Source \\
AWS Kinesis \cite{amazonrecords} & Record/Stream & Real-time Latency & Yes (AWS Only) & AWS Cloud & Proprietary \\
JetStream \cite{jetstream} & Record/Stream & Real-time Latency & Yes (Multi-Cloud) & On-Prem, Cloud & Research Prototype \\
\midrule
\multicolumn{6}{c}{\textit{Unified Data Movement}} \\
\textbf{SkyHOST (This work)} & \textbf{Hybrid Object/Stream} & \textbf{Bulk + Real-time} & \textbf{Yes (Multi-Cloud)} & \textbf{Multi-Cloud} & \textbf{Open Source} \\
\bottomrule
\end{tabular}%
}
\end{table*}

\section{RELATED WORK}
\label{sec:related}

Big data and AI workloads have fundamentally transformed data movement requirements in multi-cloud environments. Organizations handle high-throughput bulk transfers for massive datasets while simultaneously supporting low-latency streams \cite{multicloud_overview},\cite{datasource_systematic},\cite{hybrid_cloud}. These diverse datasets are spread across hybrid cloud environments and require different compute resources and transfer capabilities. However, specialized tools are typically designed to handle either stream or bulk transfers, but not both within a single framework. This forces users to deploy and manage separate infrastructure and transfer tools for each use case which results in operational overhead and complex pipeline integration. We organize existing systems into two primary classes based on their data model and optimization goals, followed by a review of unified and hybrid approaches.

\subsection{HIGH-THROUGHPUT BULK TRANSFER SYSTEMS
}

These bulk transfer systems focus primarily on moving large files and datasets efficiently. Skyplane \cite{jain2023skyplane} performs cross-cloud object transfers through gateway placement and overlay network planning. Similarly, Blaze \cite{Blaze} presents a high-performance framework built on Apache Airavata MFT to optimize the inter-cloud data movement using a single-agent architecture with parallel TCP connections and chunking. However, these systems are designed for chunk-based object-to-object transfers and lack the support for continuous ingestion model required for streaming workloads. Traditional tools like GridFTP \cite{globus}, FDT \cite{fdt} and Rucio \cite{rucio} provide capabilities for large file transfer operations but lack support for continuous low-latency data streams.

\subsection{LOW-LATENCY STREAMING SYSTEMS
}

Streaming systems prioritize real-time movement across heterogeneous environments. Platforms like Apache Kafka \cite{kafka} and Amazon Kinesis \cite{amazonrecords} provide a low-latency publish/subscribe (pub/sub) model for data streams. Tools such as MirrorMaker \cite{kafka-mirrormaker2} and Confluent Replicator \cite{confluent-replicator} enable cross-cluster data replication, while Kafka Connect \cite{kafka-connect} and its ecosystem of connectors (e.g., S3 Source connectors) facilitate integrations between streams and object stores. Streaming prototypes like JetStream \cite{jetstream} provide multi-site cloud transfer capability using adaptive batch size based on latency and cost. However, these tools excel at per-record low-latency ingestion but they are not optimized for large-scale bulk data transfers \cite{record-aware}.

Table \ref{tab:taxonomy_comparison} provides a comparative analysis of these platforms in key architectural dimensions including data model, optimization target, cross-cloud transfer, deployment, and availability.

\subsection{UNIFIED AND HYBRID APPROACHES
}

Several industrial and research systems have been proposed
to support event streaming and bulk data transfer. Existing
work typically focuses on specific aspects such as data ingestion, domain-specific architecture, bulk transfer, and stream
optimization. However, the growing diversity of modern data source formats, and deployment environments makes it difficult to achieve both interoperability and flexibility without
a unified architecture. Gobblin \cite{gobblin} (developed at LinkedIn) provides a unified framework for ingesting data from various sources (Kafka, FTP, databases) into Hadoop, managing work for both batch and near-real-time data. Similarly, Marmaray \cite{marmaray} (developed at Uber) offers a generic data ingestion framework for the Hadoop ecosystem. Liquid \cite{liquid} introduced dynamic switching of transfer protocols based on file size and network conditions. Liquid routes small messages through low-latency channels while accumulating large datasets for high-throughput channels like GridFTP. Although these systems provide unified data ingestion within their ecosystems, they lack native support for unified object stores and streaming platforms.

SnappyData \cite{snappydata} presented a unified cluster to manage streaming, transactions, and analysis using a hybrid engine with Apache Spark and GemFire to optimize streams and stored datasets. Similarly, Fannouch et al. \cite{unified_framework} proposed a Unified Data Framework (UDF) for data management, transfer, and provisioning. These systems excel at unifying data processing but still depend on specialized data movement systems.

Domain-specific research has highlighted the critical need for unified data ingestion. In healthcare, Mavrogiorgou et al. \cite{batch_HHR} built a unified data ingestion system that combines batch historical Electronic Health Records (EHRs) with real-time streaming data from Internet of Medical Things (IoMT) devices. However, it focuses on health-specific fields rather than general-purpose data movement. In multi-cloud environments, Vergilio et al. \cite{vergilio2023unified} proposed a unified reference architecture (MC-BDP) for stream processing across clouds, but their work targets the application layer rather than data movement layer.

Recent research has introduced methods to optimize systems within the streaming paradigm. KerA \cite{kera} improves ingestion throughput with dynamic partitioning, JANUS \cite{janus} reduces latency for edge IoT streams, and zStream \cite{zstream} uses adaptive micro-batching to reduce tail latency. While these systems advance the state-of-the-art in streaming, they operate within the stream paradigm and do not address the fundamental challenge of unifying bulk and stream transfers.

\begin{figure*}[ht]
    \centering
    \includegraphics[width=0.95\textwidth, height=7.5cm]{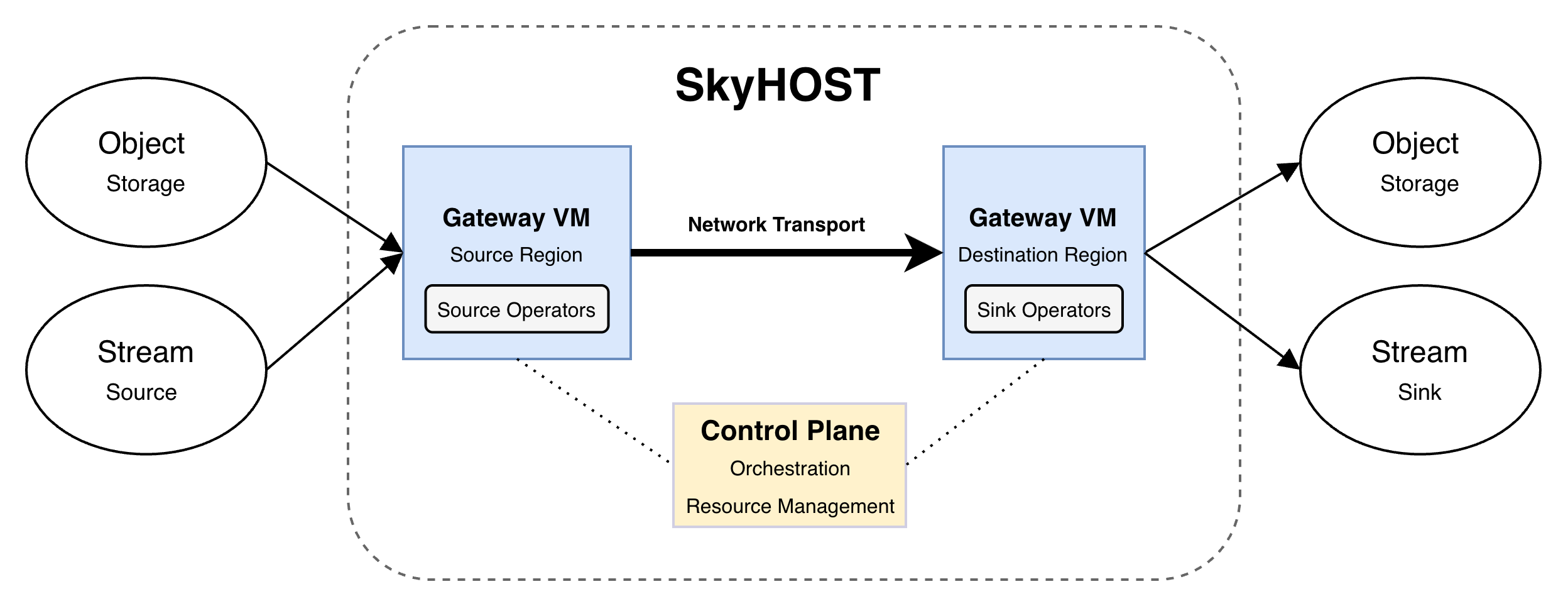}
    \caption{SkyHOST Unified System Architecture for Hybrid Object and Stream Transfer}
    \label{fig:unified_arch}
\end{figure*}

\subsection{PRIOR EXPERIMENTAL FINDINGS
}

We conducted an experimental evaluation comparing specialized tools for data transfer across various data sizes in our previous work \cite{arslan}. We selected Apache Kafka for stream ingestion and Rucio for file transfer. Experiments were carried out by transferring 10 GB datasets with record sizes ranging from small (KB) to medium (MB).

Our results demonstrated clear performance trade-offs between Apache Kafka and Rucio. Apache Kafka outperformed Rucio when working with smaller records under 1 MB for both read and write operations. However, Rucio showed better performance as the record sizes increased from 1 MB to 10 MB. These findings confirmed that no single specialized tool covers all cases, while highlighting the need for a unified approach to handle diverse data sources.

\subsection{RESEARCH GAP AND PROBLEM STATEMENT
}

Our experimental results and analysis of existing state-of-the-art systems highlight a significant research gap. Current specialized transfer tools are focused on their specific domains but unable to bridge the gap between bulk and stream transfer.
We identify the key limitations in current systems as follows:

\textit{Operational Complexity:} The lack of a unified framework increases the operational and deployment complexity for separate data transfer tools. This complexity includes handling multiple APIs, system coordination, tool configuration, parameter fine-tuning, and managing diverse data formats.

\textit{Data Model Incompatibility:} Bulk data transfer tools rely on a chunk-based data model for large binary objects, whereas streaming systems use record-aware models for structured data such as CSV or JSON. Existing systems do not bridge these hybrid data models, making it difficult to handle different data types within a single pipeline.

\textit{Limited Transfer Adaptability and Flow Control:} Current systems lack transfer adaptability based on workload characteristics or application requirements. They provide limited support to fine-tune transfer or set configurable triggers based on data formats to balance latency and throughput. Furthermore, it is difficult to manage flow control for heterogeneous pipelines with varying data arrival rates. When fast data sources connect to slower sinks, it can lead to buffer overflow, memory exhaustion, and system instability.

Based on the system limitations discussed above, we define our research question as follows: \textbf{How can we design a unified data movement framework that manages both high-throughput bulk transfers and streaming workloads via a single control plane, while eliminating the operational complexity in managing separate, specialized systems?}

\begin{table}[t]
\centering
\caption{Comparison: Specialized vs. Unified}
\label{tab:arch-comparison}
\resizebox{\columnwidth}{!}{%
\begin{tabular}{@{} lccc @{}}
\toprule
\textbf{Feature} & \textbf{Bulk Transfer} & \textbf{Stream Replication} & \textbf{SkyHOST} \\
& \textit{(e.g., Skyplane)} & \textit{(e.g., Replicator)} & \textit{(Unified)} \\
\midrule
\multicolumn{4}{c}{\textit{Native Support}} \\
Object-to-Object & \checkmark & $\times$ & \textbf{\checkmark} \\
Stream-to-Stream & $\times$ & \checkmark & \textbf{\checkmark} \\
Object-to-Stream & $\times$ & Via Connectors & \textbf{\checkmark} \\
\midrule
\multicolumn{4}{c}{\textit{Operational Complexity}} \\
Systems Req. & 1 (bulk only) & 2 (stream+conn) & \textbf{1 (unified)} \\
Config Points & Transfer-specific & Stream+Conn & \textbf{Unified} \\
Deployment & Ephemeral & Persistent & \textbf{Ephemeral} \\
\midrule
\multicolumn{4}{c}{\textit{Optimization Approach}} \\
Model & Throughput & Latency & \textbf{Adaptive} \\
Batching & Fixed Chunks & Producer Config & \textbf{Context-Aware} \\
\bottomrule
\end{tabular}%
}
\end{table}

\begin{figure*}[ht]
    \centering
    \includegraphics[width=0.8\textwidth, height=10cm]{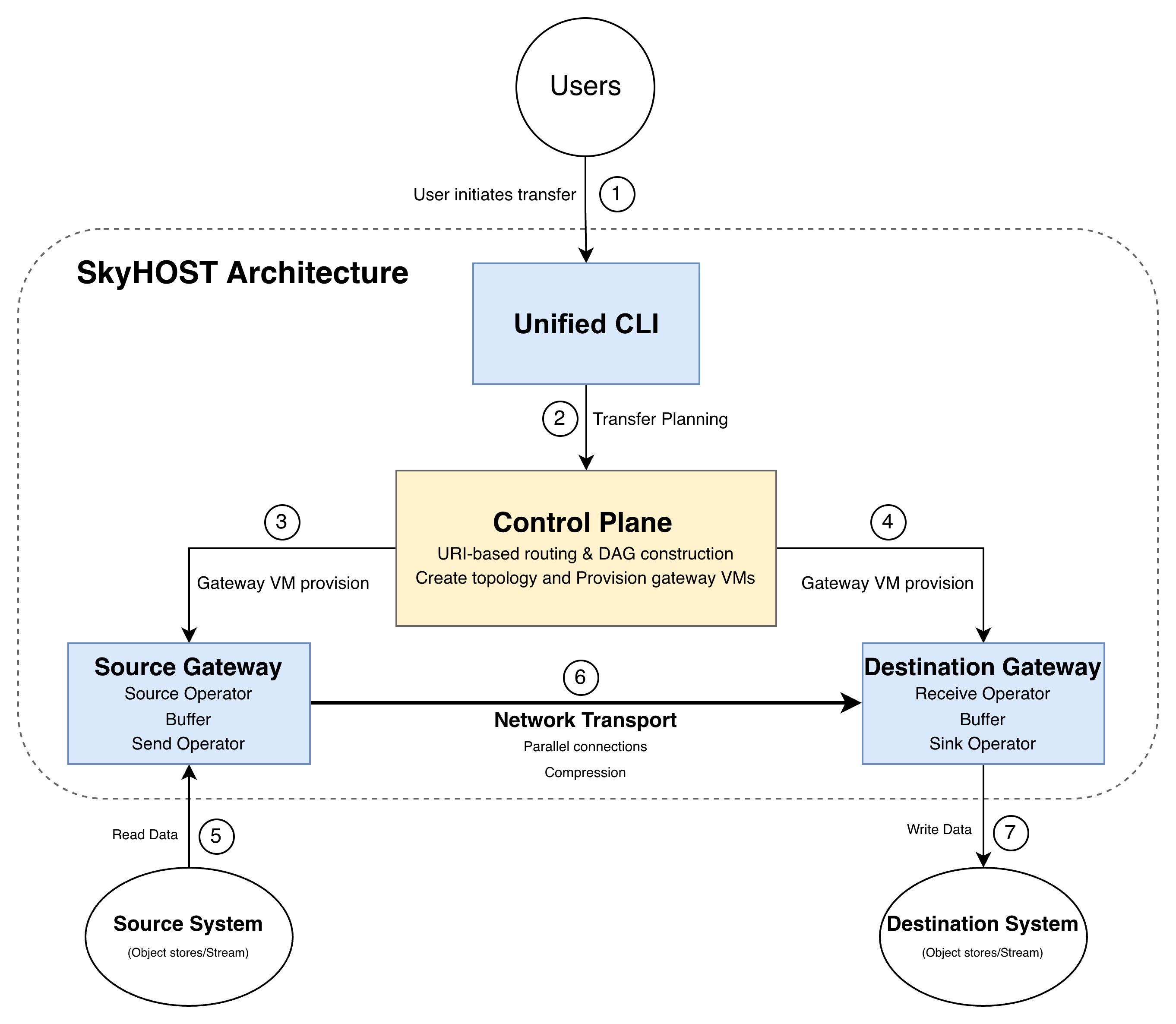}
    \caption{SkyHOST end-to-end data flow to support object-to-stream and stream-to-stream transfers}
    \label{end-to-end-dataflow}
\end{figure*}

\section{ARCHITECTURE OVERVIEW}
\label{sec:Architecture}

SkyHOST addresses the challenge of integrating high-throughput bulk transfer and continuous stream replication within a unified architecture. We built SkyHOST as an extension of the Skyplane framework \cite{jain2023skyplane} because it provides a foundation for cross-cloud object movement and overlay network architecture. We transform bulk transfer into a hybrid pipeline capable of handling real-time streaming by extending Skyplane's DAG-based operator model. SkyHOST utilizes Skyplane's gateway VM capabilities to manage cross-cloud connectivity while adding native streaming support. The system is organized as a Directed Acyclic Graph (DAG) of operators, where each stage of data ingestion, transfer, or network transit is an independent and interchangeable component. These components are executed on gateway VMs provisioned across clouds, as shown in Figure \ref{fig:unified_arch}.  

Our unified architecture manages bulk and streaming workloads under a single control plane and CLI while supporting cross-cloud transfer. The flexibility of this DAG-based design allows it to handle diverse data movement patterns, ranging from massive terabyte historical transfers to stream replication. This significantly reduces operational overhead by eliminating the requirement for managing separate tools for each data pattern and fine-tuning configurations while utilizing gateway VMs and network bandwidth across all transfer jobs.

Table \ref{tab:arch-comparison} summarizes the architectural capabilities of SkyHOST against the state-of-the-art specialized tools. The comparison highlights that specialized tools excel in one domain, whereas the SkyHOST unified architecture integrates bulk and streaming workloads into a single platform. SkyHOST provides native support for streaming workloads while reducing operational complexity and streamlining the handling of heterogeneous data sources.

SkyHOST introduces object and stream operators that support two data movement modes while maintaining a unified control plane and cross-cloud transfers. In this work, we focus on Object-to-Stream and Stream-to-Stream transfers. Stream-to-Object transfers are outside the scope of this work.

\textit{Object-to-Stream Transfer:} A format-aware source operator parses record-aware batches for structured inputs (CSV, JSON) or transfers byte-sliced micro-batches for unstructured/binary data.

\textit{Stream-to-Stream Replication:} Stream replication operators consume messages from the source topic and aggregate them into batches using configurable triggers. Batches are transmitted across regions and produced to the destination topic.

\subsection{ARCHITECTURAL COMPONENTS AND DATA FLOW}

The system consists of a control plane and a data plane that together provide end-to-end unified transfer capabilities, as shown in Figure \ref{end-to-end-dataflow}.

\subsubsection{CONTROL PLANE}
 
The SkyHOST control plane extends Skyplane's orchestration engine to manage the deployment and lifecycle of gateway VMs in specified source and destination cloud regions. It manages authentication, resource management, and cross-cloud configuration providing a unified management interface for all data movement patterns. 

\subsubsection{DATA PLANE}
The data plane executes the DAG operators on provisioned gateway VMs to perform cross-cloud transfer.

\textit{Source Gateway (SGW):} The SGW acts as the ingestion point for transfer as it executes the specific \textit{Source Operator} designed for the input type. The \textit{Object Source Operator} reads objects from storage and chunks data into batches for network transport, while the \textit{Stream Source Operator} consumes messages from the source topic and aggregates them into micro-batches based on configurable triggers.

\textit{Network Transport:} Data is transferred over TCP connections established directly between source and destination gateways. The transport layer provides the parallel connections (one per sender worker) and optional compression.

\textit{Destination Gateway (DGW):} The DGW receives incoming batches from the network and executes the \textit{Sink Operator} based on the destination type. The \textit{Stream Sink Operator} processes data chunks and produces records to the destination topic.

\subsection{KEY DESIGN PRINCIPLES AND FEATURES}

SkyHOST's unified approach addresses challenges in distributed data movement systems through the following design principles. 

\subsubsection{UNIFIED CLI AND CONTROL PLANE}

The system provides a unified CLI and control plane for all data movement tasks while eliminating the operational complexity of managing separate tools and configurations. 

\subsubsection{FORMAT-AWARE DATA TRANSFER}

SkyHOST’s architecture bridges the data model incompatibility between bulk transfer and streaming systems. The architecture supports structured data formats (CSV, JSON) for record-level processing and raw-byte transfer for unstructured or binary workloads. 

\subsubsection{BACKPRESSURE MANAGEMENT}

The system manages data flow through bounded queues that connect the operators. When the buffer hits its maximum capacity, the queue blocks the pipeline. This backpressure mechanism prevents memory exhaustion and ensures a streamlined pipeline with varying data arrival rates.

\subsubsection{CONFIGURABLE MICRO-BATCHING}

SkyHOST implements a micro-batching mechanism to balance throughput and latency, as individual records are too small to efficiently utilize available bandwidth.

The framework provides three configurable trigger types that enable automatic adaptation based on different workloads. \textit{Size-based triggers} initiate transfer when a batch reaches a specific size threshold to maximize network utilization. \textit{Time-based triggers} enforce a strict time limit to ensure messages are delivered within bounded latency preventing excessive batching delays. \textit{Count-based triggers} use a configurable message count threshold to avoid memory exhaustion during high-volume bursts. 

These configurable triggers allow the system to adapt automatically. When data arrive at higher rates, size-based batching is used for maximum throughput, while slower messages rely on time-based triggers for low latency.

\section{PERFORMANCE MODELING AND ANALYTICAL FRAMEWORK}
\label{sec:Performance}

In this section, we present the performance model to analyze the relationship between configurable parameters and performance metrics in our unified architecture. The goal is to understand performance trade-offs and parameter selection for different workloads. 

\subsection{SYSTEM NOTATIONS AND PARAMETERS}

We define our system notations using the global parameters listed in Table \ref{tab:notation}. These are categorized by system constraints, workload characteristics, and configuration parameters.

\begin{table}[t]
\centering
\caption{Summary of Analytical Model Parameters}
\label{tab:notation}
\renewcommand{\arraystretch}{1.3}
\begin{tabularx}{\columnwidth}{@{} l X @{}}
\toprule
\textbf{Symbol} & \textbf{Description [Unit]} \\
\midrule
\multicolumn{2}{l}{\textit{System Constants}} \\
$B_w$ & Network bandwidth [MB/sec] \\
$\tau$ & Per-byte processing cost for bulk read [sec/byte] \\
$T_{api}$ & Fixed object storage API overhead [sec] \\
\midrule
\multicolumn{2}{l}{\textit{Workload Characteristics}} \\
$\lambda$ & Message arrival rate [msg/sec] \\
$M_s$ & Average message size [bytes] \\
\midrule
\multicolumn{2}{l}{\textit{Configuration Parameters}} \\
$S_b$ & Target batch size [bytes] \\
$T_{max}$ & Maximum batching time [sec] \\
$C_{max}$ & Maximum message count per batch [count] \\
$S_c$ & Chunk size for object store reads [bytes] \\
$P$ & Number of partitions [count] \\
\bottomrule
\end{tabularx}
\end{table}

\subsection{MODEL ASSUMPTIONS}

\textit{Stream Replication:} We assume uniform message arrival within each batch cycle. In our experiments, $C_{max}$ and $T_{max}$ are set large so that the size trigger always fires as model validation focuses on size-based trigger.

\textit{Bulk Transfer:} We use a linear cost model for chunk processing, separating fixed API overhead ($T_{api}$) from variable per-byte costs ($\tau$). We assume pipeline parallelism across multiple chunks allows throughput to approach the bottleneck stage capacity.

\subsection{STREAM REPLICATION MODEL}

Message arrival rates and batching drive the stream replication model. SkyHOST's decoupled architecture allows batching and network transfer in a concurrent pipeline. The effective throughput $\Theta_{stream}$ is determined by the bottleneck stage of the pipeline:

\vspace{-2mm}
\begin{equation}
\Theta_{stream} = \frac{S_{b}}{\max(T_{batch}, T_{transmit})}
\label{eq:throughput-stream}
\end{equation}
\vspace{-2mm}

$\Theta_{stream}$ is measured in bytes/sec. We assume processing is overlapped with batching/transfer and not the bottleneck in cross‑region settings. Where $T_{batch}$ is determined by the first active trigger:

\begin{equation}
T_{batch} = \min\left(\frac{S_{b}}{\lambda \cdot M_s}, \quad \frac{C_{max}}{\lambda}, \quad T_{max}\right)
\end{equation}

And the network transmission time $T_{transmit}$ is: 

\begin{equation}
T_{transmit} = \frac{S_{b}}{B_w} 
\end{equation}

This throughput model validates the behavior that when network is slow $T_{transmit} > T_{batch}$, throughput is limited by bandwidth $B_w$, while if $T_{batch} > T_{transmit}$, throughput is limited by message arrival rate $\lambda \cdot M_s$. 
 
\subsection{BULK OBJECT TRANSFER MODEL}

Bulk transfer performance depends upon object storage API overhead and available bandwidth. For bulk transfer, performance is constrained by efficiency of reading and processing chunks. We model the time to process a single chunk, $T_{chunk}$, using a linear cost model:

\begin{equation}
T_{chunk} = T_{api} + \tau \cdot S_c
\label{eq:chunk-time}
\end{equation}

where $T_{api}$ captures fixed API overhead (S3 GET request, authentication, batch setup) and $\tau$ represents per-byte processing cost. 

With $P$ parallel workers, the aggregate throughput is bounded by either processing capacity or network bandwidth:

\begin{equation}
\Theta_{object} = \min\left(B_w, \quad \frac{P \cdot S_c}{T_{api} + \tau \cdot S_c}\right)
\label{eq:throughput-object}
\end{equation}

We assume $P$ parallel workers operate independently and share the same network bottleneck. For small chunks, the fixed overhead $T_{api}$ slows the transfer process. As $S_c$ increases, this overhead becomes negligible, and throughput approaches the bandwidth limit $B_w$. 

\section{SYSTEM IMPLEMENTATION}
\label{sec:System}

We implement the SkyHOST unified data movement architecture by extending the Skyplane framework, a high-performance bulk transfer system. Our extension adds native support for streaming operators and object operators, enabling unified data movement across streaming platforms and cloud object storage within a single execution framework.

\subsection{URI-BASED ROUTING AND CONTROL PLANE}

SkyHOST’s control plane parses source and destination URIs provided through the unified CLI and automatically constructs the appropriate DAG pipeline. Object store URIs (\texttt{s3://}, \texttt{gs://}, \texttt{azure://}) invoke object operators, stream URIs (\texttt{kafka://}) select streaming operators, and object store to stream (\texttt{s3://} $\rightarrow$ \texttt{kafka://}) builds a hybrid pipeline using both operators. This routing mechanism eliminates the need for users to specify the transfer mode or manage separate tool configurations.

\subsection{DATA PLANE OPERATOR DESIGN}

The SkyHOST data plane extends Skyplane's operator model with new source and sink operators for streaming workloads. We implement these operators around design principles (1) format-aware ingestion where the source operator selects its transfer strategy based on data format, and (2) decoupled pipeline stages where batching, network transfer, and destination writes operate as independent concurrent stages connected through bounded queues.

\subsubsection{OBJECT-TO-STREAM IMPLEMENTATION}
The object-to-stream operator bridges the data model mismatch between chunk-based object storage and record-oriented streaming systems. At the source gateway, the \texttt{GatewayObjStoreReadOperator} reads objects and forms either record-aware batches or byte-sliced micro-batches. The \texttt{GatewaySender} transmits these batches over parallel TCP connections. On the destination side, the \texttt{GatewayReceiver} receives chunks, optionally decompresses them, and writes them to ChunkStore. Then \texttt{GatewayKafkaWriteOperator} reads these chunks and produces messages to the destination topic.

\subsubsection{STREAM-TO-STREAM IMPLEMENTATION}

The stream-to-stream operator supports cross-cluster stream replication using configurable micro-batching. At the source gateway, the stream read operator \texttt{GatewayKafkaReadOperator} consumes messages from the source topic and aggregates them into batches using the configurable triggers. This decouples the Kafka consumer from network transfers because the \texttt{GatewaySender} operator transmits batch $N$ over TCP connections, while the consumer concurrently fills batch $N{+}1$ to maximize throughput. At the destination gateway, the \texttt{GatewayKafkaWriteOperator} deserializes the batches into records and writes them to the destination topic. SkyHOST provides at-least-once delivery semantics, and preserves partition ordering when the destination topic partitions align with the source and the partition-preservation option is enabled.

\section{EVALUATION}
\label{sec:Evaluation}

This section presents the evaluation of SkyHOST through a series of experiments that measured the throughput, scalability, and performance trade-offs across heterogeneous data sources. We validated our unified architecture through a multi-source environmental monitoring use case that requires both high-throughput bulk transfer from object storage and continuous stream replication. We analyze the results for both Kafka-to-Kafka replication and S3-to-Kafka transfer, comparing SkyHOST against state-of-the-art baseline tools Confluent Replicator and S3 Source Connector to demonstrate the benefits and trade-offs of our unified approach.  

\subsection{EXPERIMENTAL SCENARIO: MULTI-SOURCE ENVIRONMENTAL MONITORING}

European environmental monitoring presents a multi-source use case for our unified data movement architecture that handles both historical satellite data and ground-based sensor data. These diverse datasets must be consolidated into a single cluster for analysis and rapid decision-making, supporting critical applications such as flood detection, air quality monitoring, and agricultural planning. This use case highlights the real-world challenges of managing diverse data sources across multi-region deployments for environmental monitoring and analytics.

\textit{Historical Archives:} The European Environment Agency (EEA) \cite{eea_website} provides historical satellite imagery datasets. We use the Copernicus ERA5-Land dataset which includes precipitation, soil moisture, and vegetation indices stored in AWS S3 as binary files. The goal is to transfer this EEA dataset from AWS S3 into a central Kafka cluster.

\textit{Ground-Based Sensor Network:} We utilize air quality sensor data from the European Environment Agency (EEA). Ground-based sensor networks continuously generate streams of environmental data. Each regional cluster aggregates its data before transmitting to a central Kafka cluster.

This scenario presents a significant challenge for data movement that requires a system that fulfills two distinct requirements simultaneously: (1) high-throughput transfer of terabytes of historical satellite data from S3 object store to a central Kafka cluster and (2) continuous replication of sensor data from multiple regional Kafka clusters to the central cluster.

\subsection{EXPERIMENTAL SETUP}

All experiments were conducted on the AWS cloud platform across regions between us-east-1 (North Virginia) and eu-central-1 (Frankfurt). We provisioned m5.4xlarge EC2 instances for all system components, i.e., Kafka brokers, SkyHOST gateways, and baseline tools. All instances were configured with kernel-level TCP network optimizations, including the BBR congestion control and increased socket buffer sizes. 

We deployed a single gateway VM at the source and destination with Kafka topics configured with a replication factor of 1. We configure batching with $S_b = 32$ MB, $T_{max} = 10$ s, and $C_{max} = 100{,}000$ messages to ensure that the size trigger always fires for consistent batch sizes across runs. The 32 MB batch size balances throughput maximization with memory constraints, which is small enough to avoid gateway buffer exhaustion. All reported results are the average of three independent runs to ensure reliability. We measure end-to-end throughput (MB/s) and message processing rate (msgs/s), as latency characterization is left for future work.

\begin{figure}[t]
\centering
\includegraphics[width=1.0\columnwidth]{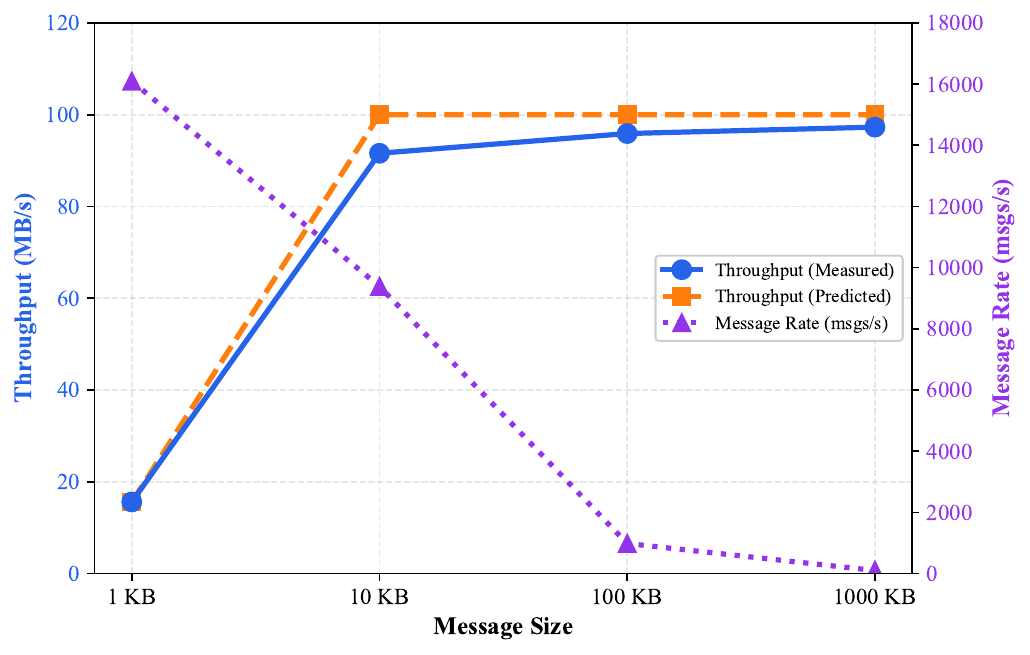}
\caption{Comparing the analytical model estimation with actual measurements in Kafka-to-Kafka replication as message size varies from 1 KB to 1000 KB}
\label{fig:kafka-to-kafka}
\end{figure}

\begin{figure}[t]
\centering
\includegraphics[width=1.0\columnwidth]{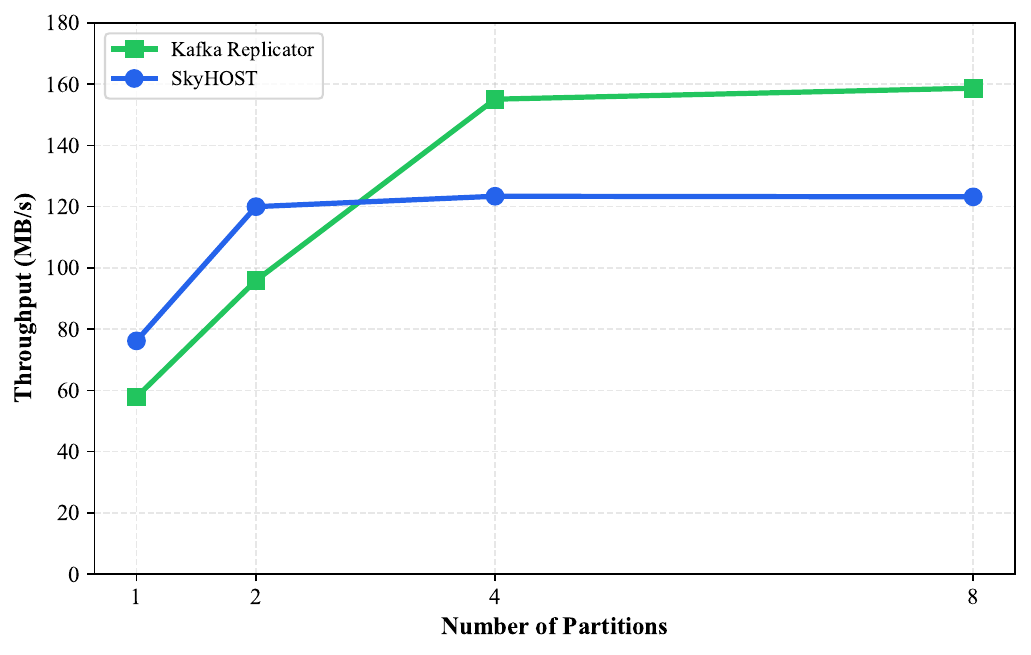}
\caption{Kafka-to-Kafka replication throughput comparison between SkyHOST and Confluent's Kafka Replicator across varying partition counts (100 KB messages, 32 MB batching)}
\label{fig:kafka-to-kafka-comparison}
\end{figure}

\begin{table}[h]
\centering
\caption{Model Parameter Values}
\label{tab:parameter-values}
\begin{tabular}{@{} l c p{4.5cm} @{}}
\toprule
\textbf{Parameter} & \textbf{Value} & \textbf{Source/Description} \\
\midrule
\multicolumn{3}{l}{\textit{Stream Replication}} \\
$B_w$ & 100 MB/s & Throughput plateau observed in Fig.~\ref{fig:kafka-to-kafka} \\
$S_b$ & 32 MB & Configured batch size \\
\midrule
\multicolumn{3}{l}{\textit{Bulk Object Transfer}} \\
$B_w$ & 140 MB/s & Throughput ceiling for chunk-based S3 reads \\
$T_{api}$ & 56 ms & Fitted using 32 MB and 64 MB data points \\
$\tau$ & 7.59 ms/MB & Fitted using 32 MB and 64 MB data points \\
$P$ & 1 & Single worker setup \\
\bottomrule
\end{tabular}
\end{table}

\subsection{RESULTS AND ANALYSIS}
We analyze the performance of SkyHOST across two primary transfer patterns i.e., Stream Replication and Bulk Object Transfer. We derive model parameters from experimental measurements and network benchmarks. Table~\ref{tab:parameter-values} reports the numerical values used for throughput model validation in Section~\ref{sec:Evaluation}. The effective bandwidth $B_w$ differs between transfer modes i.e., stream replication yields $B_w = 100$~MB/s derived from the throughput plateau in Fig.~\ref{fig:kafka-to-kafka}, while bulk transfer achieves $B_w = 140$~MB/s as chunk-based reads bypass per-record serialization. $T_{api}$ and $\tau$ are fitted from 32/64~MB data points via linear regression (Eq.~\ref{eq:chunk-time}).

\subsubsection{KAFKA-TO-KAFKA REPLICATION}

We first evaluate the throughput and message-rate trade-off for Kafka-to-Kafka replication within this unified framework. We establish a baseline for Kafka-to-Kafka performance across different message sizes (1 KB, 10 KB, 100 KB, and 1000 KB) using a fixed configuration of one partition and 32 MB inter-gateway batching. The results in Figure \ref{fig:kafka-to-kafka} show the fundamental throughput and message-rate trade-off. As the message size increases from 1 KB to 1000 KB, the overall throughput (MB/s) improves significantly while the message processing rate (msgs/s) decreases. Smaller messages (1 KB) achieve high message rates but low aggregate throughput due to per-message overhead, while larger messages achieve higher bandwidth utilization but at lower message rate. The analytical model Equation \ref{eq:throughput-stream} captures this behavior, achieving 4.1\% average prediction error under our configuration. For large messages ($\geq 10$ KB), network transmission time dominates ($T_{transmit} > T_{batch}$) and throughput approaches the effective bandwidth $B_w \approx 100$ MB/s independent of the message arrival rate. For small messages (1 KB), batching time dominates ($T_{batch} > T_{transmit}$) and throughput equals the source arrival rate. The arrival rate at 1 KB data size was $\lambda \approx 16{,}000$~msgs/s. The model accurately captures the transition from source-limited to network-limited behavior as message size increases with higher error at the smallest message size reflecting the dependency on workload-specific arrival rates in the source-limited case.

We compared our system against Confluent's Kafka Replicator, a specialized stream replication tool. This experiment measured throughput using 100 KB messages from a 1 GB EEA dataset transferred between AWS regions (us-east-1 to eu-central-1). Both systems used matched producer settings (acks=1, batch=32MB, linger=100ms, idempotence disabled), identical broker limits, and scaled concurrency with partition count for both systems (SkyHOST send-connections = partitions and Replicator tasks.max = partitions). The Replicator worker ran in the destination region, while SkyHOST used one gateway per region.

As shown in Figure \ref{fig:kafka-to-kafka-comparison}, SkyHOST achieves 76-123 MB/s while the Replicator achieves 58-159 MB/s. At low partition counts (1-2), SkyHOST outperforms Replicator by up to 32\% due to pipeline decoupling, while the destination gateway produces batch N locally to Kafka, the source gateway concurrently consumes and transmits batch N+1 over TCP connections to the destination. This decouples consumer, transfer, and producer into concurrent stages, reducing the impact of WAN round-trip latency. SkyHOST's throughput plateaus at approximately 123 MB/s for partition counts $\geq 4$, indicating a bottleneck in the single-gateway architecture that the analytical model does not capture. However, at higher partition counts Replicator achieves 29\% higher throughput, showing that its native Kafka integration enables better scaling. 

\begin{figure}[t]
\centering
\includegraphics[width=1.0\columnwidth]{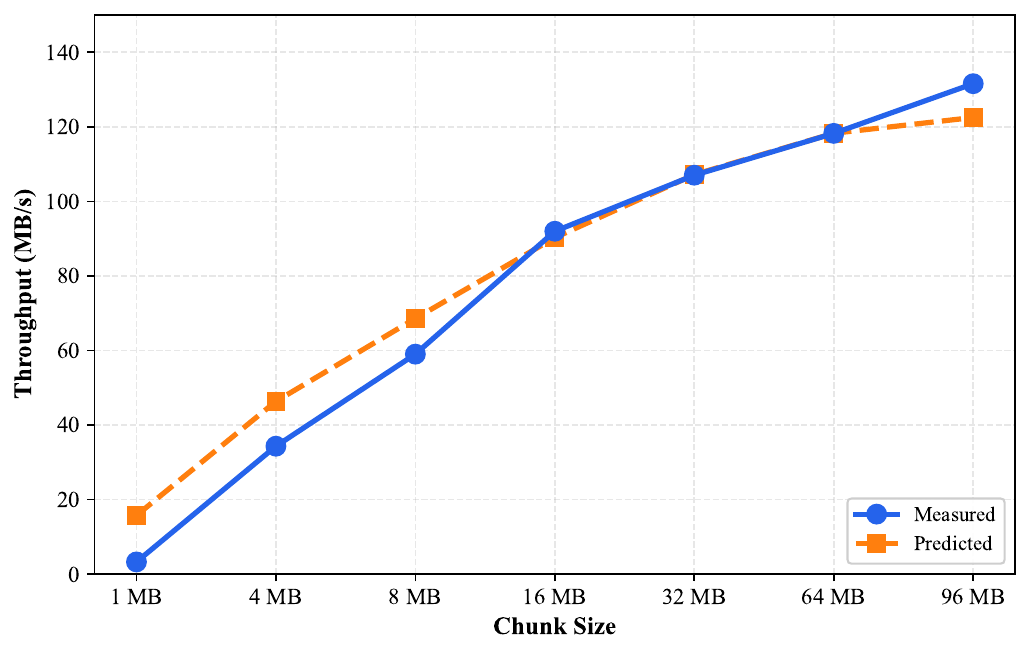}
\caption{Comparing the estimation for the analytical model with actual measurements in S3-to-Kafka transfer as chunk size varies from 1 MB to 96 MB}
\label{fig:S3-to-kafka}
\end{figure}

\begin{figure}[t]
\centering
\includegraphics[width=1.0\columnwidth]{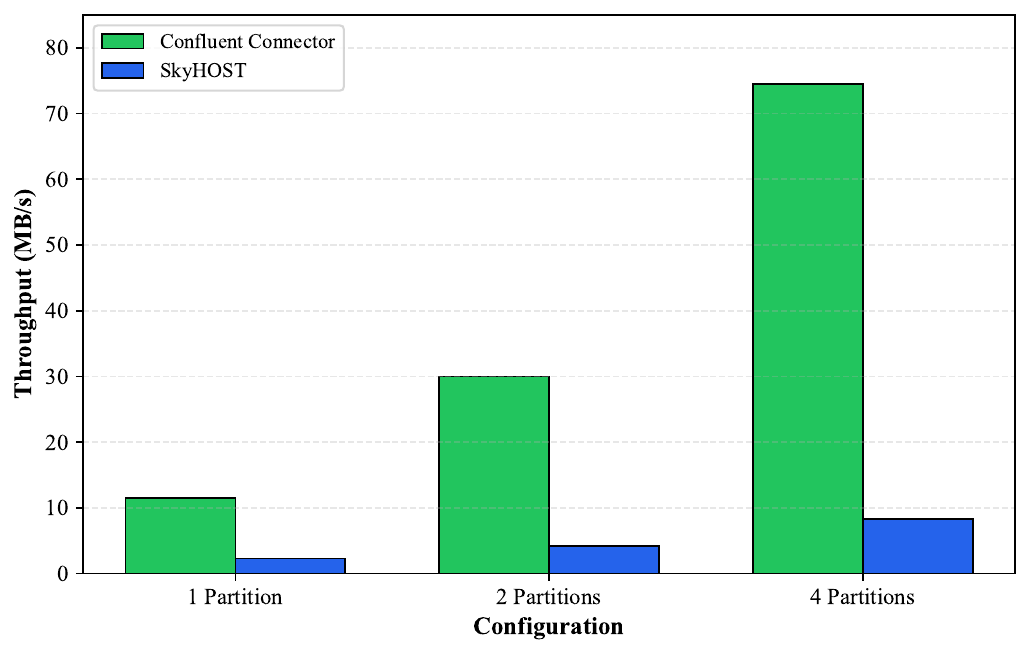}
\caption{S3-to-Kafka per-record transfer performance comparison between SkyHOST and Confluent's S3 Source Connector across varying partition counts}
\label{fig:S3-to-kafka-comparison}
\end{figure}

\subsubsection{S3-TO-KAFKA TRANSFER}

We evaluated chunk-based transfer using a 2 GB binary dataset from AWS S3 (eu-central-1) to Kafka Cluster (us-east-1) for bulk object transfer scenarios like satellite imagery. Our primary goal is high-throughput bulk transfer of large objects into Kafka. We focus on raw transfer mode because it reads objects with fixed-size range requests, slices them into chunks, and transfers them over TCP connections.

Figure \ref{fig:S3-to-kafka} shows that increasing chunk size significantly improves throughput. Larger chunks like 64 MB achieve higher throughput than smaller 1 MB chunks. This improvement validates that performance with smaller chunks is limited by the high per-request overhead of object stores, which includes S3 GET requests and network round-trips. We fit the model parameters $T_{api}$ and $\tau$ using the 32 MB and 64 MB measurements, yielding $T_{api} = 56$ ms and $\tau = 7.59$ ms/MB. The analytical model in Equations \ref{eq:chunk-time} and \ref{eq:throughput-object} achieves 2.2\% average prediction error for chunk sizes $(\geq 16\,\text{MB})$, because larger chunk sizes reduce serialization overheads. However, for smaller chunk sizes, the model shows higher prediction error due to per-record overhead and fixed API costs.

To understand the trade-offs for supporting record-aware transfer in our unified architecture, we compare SkyHOST's record-aware mode against Confluent's specialized S3 Source Connector. Figure \ref{fig:S3-to-kafka-comparison} shows the results across varying partition counts. The purpose-built Confluent connector achieves 11.5-74.5 MB/s scaling with partition count, while SkyHOST per-record mode achieves 2.3-8.3 MB/s. 

The performance difference between SkyHOST and Confluent's connector reflects a core architectural trade-off. SkyHOST's data plane is built for high-throughput bulk transfers where large objects are sliced into fixed-size blocks and transferred with minimal overhead. On the other hand, Confluent's connector is purpose-built for S3-to-Kafka with deep integration into the Kafka ecosystem for optimized record-level ingestion. 

SkyHOST provides operational simplicity and the benefits of unification detailed in Table \ref{tab:arch-comparison}. In our environmental monitoring scenario, the baseline approach required deploying separate connector instances for S3 transfer and Kafka replication. This involved VM provisioning, Docker container deployments, manually creating Kafka Connect internal topics, and handling connector-specific JSON configurations via REST API. SkyHOST provides a unified framework that consolidates these diverse data movement patterns under a single CLI where the system automatically provisions gateway VMs and initiates the transfer. This eliminates per-connector VM provisioning, Docker management, and separate configuration files, which significantly reduces deployment and operational complexity.

\section{CONCLUSION AND FUTURE WORK}
\label{sec:Conclusion}

This paper presented SkyHOST, a unified data movement architecture that simplifies cross-cloud data transfer through a single control plane and command-line interface. By extending the Skyplane framework with native support for streaming operators, SkyHOST enables unified management of heterogeneous data movement patterns across cloud object stores and streaming platforms.

Our experimental evaluation demonstrates that SkyHOST effectively reduces the operational complexity in hybrid cloud deployments. SkyHOST achieves 76-123 MB/s when replicating Kafka streams with 4.1\% average model prediction error across message sizes. For bulk S3-to-Kafka transfers, we measured 131.6 MB/s for 96 MB chunks with 2.2\% average model prediction error for chunk sizes $(\geq 16\,\text{MB})$. We further validated the practicality of the proposed architecture through a multi-source environmental monitoring use case that integrates historical satellite data with continuous sensor streams across multiple AWS regions. 

While SkyHOST's data plane is optimized for chunk-based bulk transfer, it currently achieves lower performance than specialized systems for record-aware ingestion. Addressing this limitation is a key direction for future work, including the integration of more efficient, format-specific parsing libraries into the architecture's data plane. We also plan to extend SkyHOST to support the Stream-to-Object transfer pattern and to integrate overlay network routing to minimize both transfer latency and cost. Finally, we aim to expand SkyHOST as an open-source platform to foster adoption and support sustainability-focused applications within the research and scientific community.

\bibliographystyle{IEEEtran}
\bibliography{control,bibliography}

@inproceedings{jain2023skyplane,
  title={Skyplane: Optimizing transfer cost and throughput using cloud-aware overlays},
  author={P. Jain and S. Kumar and S. Wooders and S. G. Patil and J. E. Gonzalez and I. Stoica},
  booktitle={Proc. 20th USENIX Symp. Netw. Syst. Des. Implement. (NSDI)},
  pages={1375--1389},
  year={2023}
}

@misc{amazonrecords,
  author={M. R. Sayem},
  title={Processing large records with {Amazon Kinesis Data Streams}},
  howpublished={\url{https://aws.amazon.com/blogs/big-data/processing-large-records-with-amazon-kinesis-data-streams/}},
  year={2023}
}

@inproceedings{pulsar,
  author={K. Ramasamy},
  title={Unifying messaging, queuing, streaming and light weight compute for online event processing},
  booktitle={Proc. 13th ACM Int. Conf. Distrib. Event-Based Syst. (DEBS)},
  year={2019},
  note={Art. no. 5}
}

@inproceedings{kera,
  title={Kera: Scalable data ingestion for stream processing},
  author={Marcu, Ovidiu-Cristian and Costan, Alexandru and Antoniu, Gabriel and P{\'e}rez-Hern{\'a}ndez, Mar{\'\i}a and Nicolae, Bogdan and Tudoran, Radu and Bortoli, Stefano},
  booktitle={2018 IEEE 38th International Conference on Distributed Computing Systems (ICDCS)},
  pages={1480--1485},
  year={2018},
  organization={IEEE}
}

@inproceedings{arslan,
  author={M. Tariq and O. Marcu and G. Danoy and P. Bouvry},
  title={Towards unified data ingestion and transfer for the computing continuum},
  booktitle={Proc. IEEE Int. Conf. Big Data (BigData)},
  year={2023},
  month=dec,
  pages={1978--1981}
}

@article{kafka,
  title={A survey on networked data streaming with {Apache Kafka}},
  author={T. P. Raptis and A. Passarella},
  journal={IEEE Access},
  year={2023}
}

@article{jetstream,
  title={{Jetstream}: Enabling high throughput live event streaming on multi-site clouds},
  author={R. Tudoran and A. Costan and O. Nano and I. Santos and H. Soncu and G. Antoniu},
  journal={Future Gener. Comput. Syst.},
  volume={54},
  pages={274--291},
  year={2016}
}

@article{janus,
  title={{JANUS}: Latency-aware traffic scheduling for {IoT} data streaming in edge environments},
  author={Z. Wen and others},
  journal={IEEE Trans. Services Comput.},
  year={2023}
}

@article{zstream,
  title={{zStream}: Towards a low latency micro-batch streaming system},
  author={S. Lee and Y. Jeong and K. Park and G. Jung and S. Park},
  journal={Cluster Comput.},
  volume={26},
  number={5},
  pages={2773--2787},
  year={2023}
}

@article{tbtopb,
  title={Synchrotron big data science},
  author={C. Wang and U. Steiner and A. Sepe},
  journal={Small},
  volume={14},
  number={46},
  year={2018},
  note={Art. no. 1802291}
}

@article{smartcities,
  title={Goal-driven scheduling model in edge computing for smart city applications},
  author={Y. Kim and S. Park and S. Shahkarami and R. Sankaran and N. Ferrier and P. Beckman},
  journal={J. Parallel Distrib. Comput.},
  volume={167},
  pages={97--108},
  year={2022}
}

@article{tomography,
  title={{TomoGAN}: Low-dose synchrotron x-ray tomography with generative adversarial networks: Discussion},
  author={Z. Liu and T. Bicer and R. Kettimuthu and D. Gursoy and F. De Carlo and I. Foster},
  journal={J. Opt. Soc. Amer. A},
  volume={37},
  number={3},
  pages={422--434},
  year={2020}
}

@article{smartagriculture,
  title={A scalable smart farming big data platform for real-time and batch processing based on lambda architecture},
  author={M. E. M. El Aissi and S. Benjelloun and Y. Lakhrissi and S. Ali},
  journal={J. Syst. Manage. Sci.},
  volume={13},
  number={2},
  pages={17--30},
  year={2023}
}

@article{data_intensive,
  title={Distributed systems for data-intensive computing in cloud environments: A review of big data analytics and data management},
  author={Z. Arif and S. R. Zeebaree},
  journal={Indonesian J. Comput. Sci.},
  volume={13},
  number={2},
  year={2024}
}

@inproceedings{globus,
  title={The {Globus} striped {GridFTP} framework and server},
  author={W. Allcock and J. Bresnahan and R. Kettimuthu and M. Link},
  booktitle={Proc. 2005 ACM/IEEE Conf. Supercomput. (SC'05)},
  year={2005},
  note={Art. no. 54}
}

@article{record-aware,
  title={A survey on the evolution of stream processing systems},
  author={M. Fragkoulis and P. Carbone and V. Kalavri and A. Katsifodimos},
  journal={VLDB J.},
  volume={33},
  number={2},
  pages={507--541},
  year={2024}
}

@article{rucio,
  title={Rucio: Scientific data management},
  author={M. Barisits and others},
  journal={Comput. Softw. Big Sci.},
  volume={3},
  number={1},
  year={2019},
  note={Art. no. 11}
}

@inproceedings{multicloud_overview,
  title={An overview of multi-cloud computing},
  author={J. Hong and T. Dreibholz and J. A. Schenkel and J. A. Hu},
  booktitle={Proc. Workshops Int. Conf. Adv. Inf. Netw. Appl.},
  pages={1055--1068},
  year={2019}
}

@article{datasource_systematic,
  title={A systematic review of big data integration challenges and solutions for heterogeneous data sources},
  author={F. Z. Rozony and M. Aktar and M. Ashrafuzzaman and A. Islam},
  journal={Acad. J. Bus. Admin. Innov. Sustain.},
  volume={4},
  number={04},
  pages={1--18},
  year={2024}
}

@misc{kafka-connect,
  author={{Apache Kafka}},
  title={{Kafka Connect: Streaming integration made simple}},
  howpublished={\url{https://kafka.apache.org/documentation/\#connect}},
  year={2024}
}

@article{hybrid_cloud,
  title={State-of-the-art, challenges, and open issues in the integration of {Internet of Things} and cloud computing},
  author={M. D{\'\i}az and C. Mart{\'\i}n and B. Rubio},
  journal={J. Netw. Comput. Appl.},
  volume={67},
  pages={99--117},
  year={2016}
}

@misc{kafka-mirrormaker2,
  author={{Apache Kafka}},
  title={{KIP-382: MirrorMaker 2.0}},
  year={2019},
  howpublished={\url{https://cwiki.apache.org/confluence/display/KAFKA/KIP-382\%3A+MirrorMaker+2.0}}
}

@misc{confluent-replicator,
  author={{Confluent Inc.}},
  title={{Confluent Replicator}},
  year={2024},
  howpublished={\url{https://docs.confluent.io/platform/current/multi-dc-deployments/replicator/index.html}}
}

@misc{fdt,
  author={{CERN/MonALISA}},
  title={{FDT: Fast Data Transfer}},
  year={2024},
  howpublished={\url{https://monalisa.cern.ch/FDT/}}
}

@article{health1,
  title={The use of big data analytics in healthcare},
  author={K. Batko and A. {\'S}l{\k{e}}zak},
  journal={J. Big Data},
  volume={9},
  number={1},
  year={2022},
  note={Art. no. 3}
}

@article{health2,
  title={A comprehensive survey on machine learning-based big data analytics for {IoT}-enabled smart healthcare system},
  author={W. Li and others},
  journal={Mobile Netw. Appl.},
  volume={26},
  pages={234--252},
  year={2021}
}

@article{health3,
  title={A task-level adaptive {MapReduce} framework for real-time streaming data in healthcare applications},
  author={F. Zhang and J. Cao and S. U. Khan and K. Li and K. Hwang},
  journal={Future Gener. Comput. Syst.},
  volume={43},
  pages={149--160},
  year={2015}
}

@article{IoT1,
  title={Identification and authentication in healthcare {Internet-of-Things} using integrated fog computing based blockchain model},
  author={S. Shukla and S. Thakur and S. Hussain and J. G. Breslin and S. M. Jameel},
  journal={Internet Things},
  volume={15},
  year={2021},
  note={Art. no. 100422}
}

@article{IoT2,
  title={An effective handling of secure data stream in {IoT}},
  author={J. Jang and I. Y. Jung and J. H. Park},
  journal={Appl. Soft Comput.},
  volume={68},
  pages={811--820},
  year={2018}
}

@article{gobblin,
  title={Gobblin: Unifying data ingestion for {Hadoop}},
  author={L. Qiao and others},
  journal={Proc. VLDB Endowment},
  volume={8},
  number={12},
  pages={1764--1769},
  year={2015}
}

@article{dataflow,
  title={The dataflow model: A practical approach to balancing correctness, latency, and cost in massive-scale, unbounded, out-of-order data processing},
  author={T. Akidau and others},
  journal={Proc. VLDB Endowment},
  volume={8},
  number={12},
  pages={1792--1803},
  year={2015}
}

@article{vergilio2023unified,
  title={A unified vendor-agnostic solution for big data stream processing in a multi-cloud environment},
  author={T. Vergilio and A.-L. Kor and D. Mullier},
  journal={Appl. Sci.},
  volume={13},
  number={23},
  year={2023},
  note={Art. no. 12635}
}

@article{batch_HHR,
  title={Batch and streaming data ingestion towards creating holistic health records},
  author={A. Mavrogiorgou and A. Kiourtis and G. Manias and C. Symvoulidis and D. Kyriazis},
  journal={Emerging Sci. J.},
  volume={7},
  number={2},
  pages={339--353},
  year={2023}
}

@inproceedings{Blaze,
  author={S. Marru and others},
  title={Blaze: A high-performance, scalable, and efficient data transfer framework with configurable and extensible features: Principles, implementation, and evaluation of a transatlantic inter-cloud data transfer case study},
  booktitle={Proc. IEEE 16th Int. Conf. Cloud Comput. (CLOUD)},
  year={2023},
  pages={58--68}
}

@misc{eea_website,
  author={{European Environment Agency}},
  title={{European Environment Agency}},
  howpublished={\url{https://www.eea.europa.eu/en}},
  note={Accessed: 2025-10-25},
  year={2025}
}

@misc{marmaray,
  title={Marmaray: An open source generic data ingestion and dispersal framework and library for {Apache Hadoop}},
  author={D. Chen and O. Joshi},
  year={2018},
  howpublished={Uber Eng. Blog},
  url={https://eng.uber.com/marmaray-hadoop-ingestion-open-source/}
}

@inproceedings{liquid,
  title={Liquid: A scalable and adaptive middleware for hybrid data transfer},
  author={T. Kosar and others},
  booktitle={Proc. IEEE Int. Symp. Cluster, Cloud Grid Comput.},
  year={2012}
}

@inproceedings{snappydata,
  title={{SnappyData}: A unified cluster for streaming, transactions and interactive analytics},
  author={B. Mozafari and others},
  booktitle={Proc. CIDR},
  volume={17},
  pages={8--11},
  year={2017}
}

@inproceedings{unified_framework,
  title={Unified data framework for enhanced data management, consumption, provisioning, processing and movement},
  author={A. Fannouch and Y. Gahi and J. Gharib},
  booktitle={Proc. 7th Int. Conf. Netw. Intell. Syst. Security},
  pages={1--7},
  year={2024}
}

\end{document}